# Improving the Accuracy and Interpretability of Neural Networks for Wind Power Forecasting

Wenlong Liao, Fernando Porté-Agel, Jiannong Fang, Birgitte Bak-Jensen, Zhe Yang, Gonghao Zhang

*Abstract*—Deep neural networks (DNNs) are receiving increasing attention in wind power forecasting due to their ability to effectively capture complex patterns in wind data. However, their forecasted errors are severely limited by the local optimal weight issue in optimization algorithms, and their forecasted behavior also lacks interpretability. To address these two challenges, this paper firstly proposes simple but effective triple optimization strategies (TriOpts) to accelerate the training process and improve the model performance of DNNs in wind power forecasting. Then, permutation feature importance (PFI) and local interpretable model-agnostic explanation (LIME) techniques are innovatively presented to interpret forecasted behaviors of DNNs, from global and instance perspectives. Simulation results show that the proposed TriOpts not only drastically improve the model generalization of DNNs for both the deterministic and probabilistic wind power forecasting, but also accelerate the training process. Besides, the proposed PFI and LIME techniques can accurately estimate the contribution of each feature to wind power forecasting, which helps to construct feature engineering and understand how to obtain forecasted values for a given sample.

## I. INTRODUCTION

IN power systems, wind power is widely recognized as an efficient renewable energy source for electricity generation. In recent decades, there has been a significant global increase in wind power integration [1]. In particular, Swiss wind power generates 5% more electricity in 2022 than last year. However, compared to conventional forms of electricity generation, wind power introduces many more uncertainties. As the penetration of wind power increases, accurate and reliable wind power forecasting is essential to ensure the safe and stable operation of power systems [2].

There is a large body of publications on wind power forecasting, and existing methods can be broadly categorized into three main groups [3]: physical models, statistical models, and machine learning models. Specifically, physical models use detailed physical parameters (such as terrain and obstacles) and numerical weather prediction (NWP) data as inputs to complex microscale computational models to simulate wind farm flows and calculate wind speeds at wind turbine hub heights [4]. Subsequently, the forecasted wind speeds are mapped onto the corresponding wind power curve, typically supplied by the manufacturer of wind turbines, to forecast wind power outputs. Although physical models can be a favorable option for medium-term and long-term wind power forecasting, they present challenges in terms of computational complexity [5].

Another popular wind power forecasting method is to input historical wind power generation values into statistical models [6], such as the autoregressive moving average model, autoregressive model, gray forecasting model, and persistence model. For example, the work in [7] uses an autoregressive integrated moving average model to capture the intermittent nature of offshore wind power. In [8], an autoregressive model is constructed to account for spatiotemporal dependencies within wind power time series data. While statistical models provide cost-effective forecasting solutions, their accuracy is often limited, especially for medium-term and long-term wind power forecasting.

In recent years, machine learning models have received significant attention in wind power forecasting due to their potential to capture complex patterns in wind power and NWP data. Machine learning models include a variety of techniques [9], ranging from the traditional support vector regression, extreme gradient boosting, linear regression, and regression tree to the more recent multi-layer perception (MLP), long short-term memory (LSTM), gated recurrent units (GRU), and temporal convolutional network (TCN) [10]. For example, the work in [11] employs the MLP to map the NWP data into wind power. To capture temporal features of wind power data, an LSTM is designed in [12]. Furthermore, the LSTM is extended into the GRU in [13], so as to improve the time efficiency of the LSTM in wind power forecasting. Generally, these models can capture the complex nonlinear relationships between NWP data, historical wind power generation values, and real values, resulting in highly reliable and accurate wind power forecasting. Compared to statistical models, these recent and advanced machine learning models, especially deep neural networks (DNNs), generally have higher accuracy, but they suffer from two limitations.

The first limitation is that the optimization algorithms used

This work is funded by the Swiss Federal Office of Energy (Grant No. SI/502135–01). Also, this work is carried out in the frame of the "UrbanTwin: An urban digital twin for climate action: Assessing policies and solutions for energy, water and infrastructure" project with the financial support of the ETH-Domain Joint Initiative program in the Strategic Area Energy, Climate and Sustainable Environment.

Wenlong Liao, Fernando Porté-Agel, and Jiannong Fang are with the Wind Engineering and Renewable Energy Laboratory, Ecole Polytechnique Federale de Lausanne (EPFL), Lausanne 1015, Switzerland.

Birgitte Bak-Jensen is with the AAU Energy, Aalborg University, Aalborg 9220, Denmark.

Zhe Yang is with Department of Electrical Engineering, The Hong Kong Polytechnic University, HongKong 999077, HongKong.

Gonghao Zhang is with the Department of Electrical and Electronic Engineering, The University of Hong Kong, Hong Kong 999077, Hong Kong.

to update the weights of DNNs have difficulties in obtaining the globally optimal weights, which limits the model generalization of DNNs [14]. Specifically, the forecasted errors of wind power and the training time of DNNs mainly depend on the used optimization algorithms [15], such as root mean square propagation (RMSprop), Nesterov-accelerated adaptive moment estimation (Nadam), adaptive moment estimation (Adam), and adaptive moment estimation with infinity norm (Adamax). The local optimal weight issue in these optimization algorithms severely affects the model performance of DNNs in wind power forecasting. Therefore, it remains a challenge to optimize the training process of DNNs in wind power forecasting.

The second limitation is that DNNs lack interpretability (i.e., it means that the user cannot know how each of the input features contributes to the forecasted values). In particular, DNNs are often viewed as black-box models, which hinder the understanding of the underlying factors influencing forecasts. This opacity can lead to reduced trust and limited adoption of DNN-based wind power forecasting in critical applications, such as power system operations. To date, limited attention has been paid to addressing the lack of interpretability associated with DNNs in wind power forecasting. While some recent publications have attempted to use Shapley values to analyze the inner workings of DNNs in solar and wind power forecasting [16], [17], it is worth noting that the computational requirements associated with Shapley value calculations can be quite substantial. Therefore, the interpretability of DNNs in wind power forecasting is another challenge.

In this context, this paper is dedicated to addressing the two above-mentioned limitations of DNNs in wind power forecasting by proposing novel techniques related to the interpretability and training process of DNNs. The main contributions are summarized as follows:
- **Optimizing Training Process**: The simple but effective triple optimization strategies (TriOpts) are proposed to accelerate the training process and improve the model performance of DNNs in wind power forecasting. Moreover, the proposed TriOpts can be easily embedded into various optimization algorithms (e.g., Adam, RMSprop, Adamax, Nadam, etc.).
- **Adding Interpretability to DNNs**: In response to the inherent lack of interpretability in DNNs, the permutation feature importance (PFI) and the local interpretable model-agnostic explanation (LIME) techniques are presented to estimate the contribution of individual input features in wind power forecasting, from global and instance perspectives.
- **Conducting Comprehensive Case Study**: A comprehensive case study is conducted on seven real datasets from three locations, which verifies the proposed techniques for improving the model performance, time efficiency, and interpretability of various DNNs.

The rest of the paper is organized as follows. Section II formulates the TriOpts to optimize the training process of DNNs. Section III formulates the LIME and PFI techniques to interpret the DNNs. In section IV, simulation and analysis are conducted on seven real-world datasets. Finally, section V presents the conclusion.

## II. OPTIMIZING TRAINING PERFORMANCE WITH THE TRIPLE OPTIMIZATION STRATEGIES

### A. Problem Statement

As shown in Fig. 1, after initializing the structure and parameters, one of the most critical steps is to update the weights of DNNs by using an optimization algorithm (e.g., Adam), which affects the generalization performance and training process of DNNs.

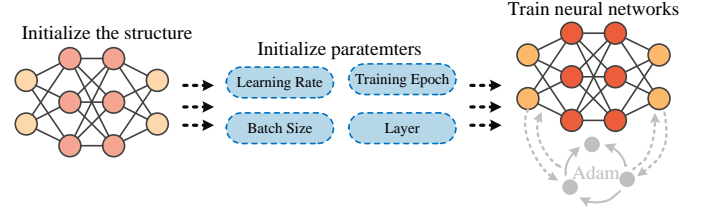

Fig. 1. The training steps of deep neural networks.

As one of the most popular and widely used optimization algorithms, the Adam algorithm has recently been broadly adopted for optimizations of DNNs in various fields [15]. Therefore, we will employ the Adam algorithm as an example to describe how the optimization algorithm updates the weights of DNNs, and other optimization algorithms can be treated in a similar way.

The core idea of the Adam algorithm is to update the weights at each time step. Unlike simple gradient descent, Adam introduces two crucial parameters $\beta_1$ and $\beta_2$ to control the weights of the first moment $m$ and second moment $v$, respectively. Typically, they are close to 1.

At the beginning of the Adam algorithm, the first moment $m$ and second moment $v$ are initialized as zero vectors. At each time step, we need to calculate the gradient $g_t$ of the loss function of model parameters (the loss function can be the mean squared error for wind power forecasting):

$$g_t = \nabla L(\theta_{t-1}) \quad (1)$$

where $\nabla L(\theta_{t-1})$ is the gradient of the loss function $L$ with respect to the parameter $\theta_{t-1}$.

Then, the parameter $\beta_1$ is used to control the weight of the first moment $m_t$ and calculate an exponentially weighted moving average of the first moment at time step $t$:

$$m_t = \beta_1 m_{t-1} + (1 - \beta_1) g_t \quad (2)$$

Eq. (2) ensures that the first moment $m$ retains information about past gradients.

Next, the parameter $\beta_2$ is used to control the weight of the second moment $v$ and calculate an exponentially weighted moving average of the second moment at time step $t$:

$$v_t = \beta_2 v_{t-1} + (1 - \beta_2) g_t^2 \quad (3)$$

Similarly, Eq. (3) ensures that the second moment $v$ retains information about past squared gradients.

Subsequently, an important correction is conducted to compensate for estimation errors at the initial time steps. This is known as bias correction. The bias-corrected first moment $\hat{m}_t$ and second moment $\hat{v}_t$ can be calculated as:

$$\hat{m}_t = \frac{m_t}{1-\beta_1^t} \tag{4}$$

$$\hat{v}_t = \frac{v_t}{1-\beta_2^t} \tag{5}$$

where $\beta_1^t$ represents $\beta_1$ to the power $t$; and $\beta_2^t$ represents $\beta_2$ to the power $t$.

Finally, the bias-corrected first and second moments can be used to update the model parameters at time step $t$:

$$\theta_t = \theta_{t-1} - \alpha \frac{\hat{m}_t}{\sqrt{\hat{v}_t}+\varepsilon} \tag{6}$$

where $\alpha$ is the learning rate; $\varepsilon$ is a small constant to prevent division by zero.

Although Eq. (6) combines information about the mean and variance of the gradients to adjust the weights of DNNs during training, it also has some issues, such as local optimal weight issue [14], [15], [18], which severely affects the model performance of DNNs. That's why we propose the TriOpts to improve the optimization algorithms.

### B. Triple Optimization Strategies

To address the limitations of the optimization algorithms mentioned above, the TriOpts are developed to improve the training process and generalization performance of DNNs. The TriOpts is a combination of strategies, which include gradient centralization, adaptive learning rates, and the addition of uniformly distributed noises.

First, we propose to apply a gradient centralization strategy to stabilize and accelerate the training process by ensuring that gradients are centered and well-behaved. This strategy leads to more consistent and predictable training dynamics, making it easier for optimization algorithms to find good weights without suffering from divergent gradients or slow convergence [18]. Specifically, the key operation of the gradient centralization strategy is to tune gradient computation and parameter updates as follows:

$$\tilde{g}_t = \nabla L(\theta_{t-1}) - \mu(\nabla L(\theta_{t-1})) \tag{7}$$

$$\mu(\nabla L(\theta_{t-1})) = \frac{1}{C}\sum_{i=1}^{C}\nabla L(\theta_{t-1})_i \tag{8}$$

where $C$ represents the number of channels or dimensions, often used in DNNs (e.g., LSTM and MLP); $\mu(\nabla L(\theta_{t-1}))$ is the mean value of gradients; $\tilde{g}_t$ is the centralized gradient, which will replace the original gradient and participate in the computation in Eq. (2) and Eq. (3).

Secondly, an adaptive learning rate is proposed to help the DNNs converge to the optimal solution and reduce the over-fitting problem. Normally, the adaptive learning rate decay strategy can modulate the learning rate during training to balance the trade-off between rapid convergence early in training and fine-tuning closer to convergence [19], ultimately improving generalization performance. Specifically, a cosine function is employed to gradually decrease the learning rate, which starts at the initial learning rate and smoothly decreases toward zero. Its formula is as follows:

$$\alpha_t = \alpha_0 \frac{1+\cos(\pi \times t/T_e)}{2} \tag{9}$$

where $\alpha_t$ is the learning rate at training epoch $t$; $\alpha_0$ is the initial learning rate; and $T_e$ is the total number of training epochs. For the optimization algorithms (e.g., Adam) without the Triopts, the fixed learning rate is empirically set to 0.001 to ensure convergence. For the updated optimization algorithms with the TriOpts, the initial learning rate is set to 0.1, because a large initial learning rate speeds up convergence.

Thirdly, to improve the robustness of DNNs, uniformly distributed noises are added to the parameter $\theta_t$. In this case, Eq. (6) will generalize into a new form:

$$\theta_t = \theta_{t-1} - \alpha \frac{\hat{m}_t}{\sqrt{\hat{v}_t}+\varepsilon} + \text{rand}(-\tau,\tau) \tag{10}$$

where $\tau$ is a noise parameter to control the noises. Normally, hyper-parameter optimization and cross-validation, as recommended in [20], can be performed on the validation set to find the appropriate noise parameter.

Obviously, the TriOpts do not require very tedious steps to modify the optimization algorithm, so they can be easily embedded into various optimization algorithms. For example, after embedding the TriOpts to the Adam algorithm, the specific steps of the updated Adam algorithm are shown in Algorithm 1. The proposed TriOpts can be embedded into other optimization algorithms (e.g., Nadam, RMSprop, and Adamax) in a similar way, and their specific steps can be found in our public repository (https://github.com/wenlongliaoEE/TriOpts/).

Generally, the proposed TriOpts offer several advantages in the training of DNNs:
- By combining gradient centralization, adaptive learning rates, and the addition of uniformly distributed noises, the proposed TriOpts help improve model performance and accelerate the training process.
- The proposed TriOpts can be used in conjunction with various optimization algorithms (e.g., Adam, Nadam, RMSprop, Adamax, etc.) to enhance their performance without requiring a complete redesign of the optimization process.

---

**Algorithm 1:** The updated Adam algorithm with the TriOpts

1 **Input** $\alpha_0$, $T_e$, $\tau$ : initial learning rate, training epoch, and noise parameter
2 **Input** $\beta_1$, $\beta_1 \in [0,1)$: parameters to control the first and second moments
3 **Input** $L(\theta)$: loss function with respect to the parameter $\theta$
4 **Input** $\theta_0$, $m_0$, $v_0$: initial parameter vector, first moment, and second moment
5  $t \leftarrow 0$ (Initialize time step)
6  **for** $t=1,2,\ldots,T_e$ **do**
7   $\quad \tilde{g}_t \leftarrow \nabla L(\theta_{t-1}) - \mu(\nabla L(\theta_{t-1}))$  Get centralized gradients at time step $t$
8   $\quad m_t \leftarrow \beta_1 m_{t-1} + (1-\beta_1)\tilde{g}_t$  Get the first moment at time step $t$
9   $\quad v_t \leftarrow \beta_2 v_{t-1} + (1-\beta_2)\tilde{g}_t^2$  Get the second moment at time step $t$
10  $\quad \hat{m}_t \leftarrow m_t/(1-\beta_1^t)$  Get the bias-corrected first moment at time step $t$
11  $\quad \hat{v}_t \leftarrow v_t/(1-\beta_2^t)$  Get the bias-corrected second moment at time step $t$
12  $\quad \alpha_t \leftarrow \alpha_0 (1+\cos(\pi \times t/T_e))/2$  Get the learning rate at time step $t$
13  $\quad \theta_t \leftarrow \theta_{t-1} - \alpha_t \hat{m}_t /(\sqrt{\hat{v}_t}+\varepsilon) + \text{rand}(-\tau,\tau)$  Update the parameters
14 **end for**
15 **Return** $\theta_t$ (Trained parameters of DNNs)

---

## III. ADDING INTERPRETABILITY TO DEEP NEURAL NETWORKS

DNNs are often viewed as black-box models, which hinder the understanding of the underlying factors influencing forecasts. By interpretability, we mean that the user can know how each of the input features contributes to the forecasted values [16], [17]. To interpret the DNNs, this section proposes to apply PFI and LIME techniques [21], [22] to estimate the contribution of individual input features in wind power forecasting, from global and instance perspectives.

### A. Global Interpretability to Deep Neural Networks

To estimate the contribution of individual input features in wind power forecasting from a global perspective, we present the PFI technique to modify input features. The framework of the PFI technique is presented in Fig. 2.

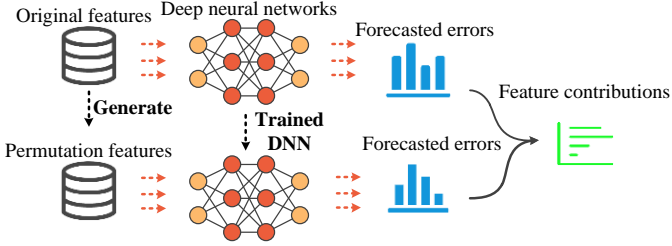

Fig. 2. The framework of the PFI technique.

Firstly, the training set is used to train a DNN (e.g., LSTM, MLP, etc.). The model performance of the DNN can be evaluated by some error metric. For example, the mean square error (MSE) is used to represent the model performance of the original input features-based DNN on the test set:

$$e_{\text{ori}} = f(Y, \text{DNN}(X)) \tag{11}$$

where $e_{\text{ori}}$ is MSE of the original input features-based DNN on the test set; $Y$ is the real wind power; $X$ denotes original input features (i.e., NWP data or historical wind power) of the DNN; DNN($X$) is the forecasting wind power; and $f$ is the MSE function.

Secondly, to effectively break the relationship between the $i^{\text{th}}$ feature and the wind power, we permute (e.g., shuffle) the values of the $i^{\text{th}}$ feature to obtain the permutation features. Then, the permutation features are fed into the pre-trained DNN to obtain the MSE on the test set:

$$e_{\text{per},i} = f(Y, \text{DNN}(X_{-i})) \tag{12}$$

$X_{-i}$ denotes permutation features (i.e., the $i^{\text{th}}$ input feature has been shuffled) of the DNN; and $e_{\text{per},i}$ is the MSE of the DNN on the test set after shuffling the $i^{\text{th}}$ input feature.

Finally, we can use the Eq. (11) and Eq. (12) to estimate the contribution of the $i^{\text{th}}$ input features in wind power forecasting from a global perspective:

$$\text{FI}_i = e_{\text{per},i} - e_{\text{ori}} \tag{13}$$

where $\text{FI}_i$ is the importance of the $i^{\text{th}}$ feature, which represents how much error can be reduced by the addition of the $i^{\text{th}}$ feature.

Obviously, the proposed PFI technique is model independent, and can be applied to explain a variety of DNNs [21]. In addition, the results of the PFI technique are easy to understand and interpret. It provides a simple metric that indicates how much each feature contributes to the model performance. This helps non-specialists to better understand the behavior of the DNN in wind power forecasting.

### B. Instance Interpretability to Deep Neural Networks

To estimate the contribution of individual input features in wind power forecasting from an instance perspective, we present the LIME technique to explain how DNNs make specific forecasts on individual instances [22]. The key idea is to understand the behavior of the DNN by generating a set of nearby approximate instances, and then constructing an interpretable model (e.g., linear regression, decision trees, etc.) to approximate the forecasted behavior of the DNN on a given specific instance. The framework of the LIME technique is shown in Fig. 3.

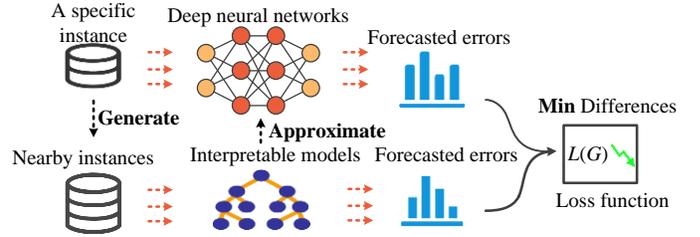

Fig. 3. The framework of the LIME technique.

Firstly, we randomly generate a set of perturbed instances $(\tilde{X}_1, \tilde{X}_2, ..., \tilde{X}_n)$ that are slightly different in feature space from the original instance $X$ to be interpreted. The DNN can be used to obtain the forecasted value for these perturbed instances $(\text{DNN}(\tilde{X}_1), \text{DNN}(\tilde{X}_2), ..., \text{DNN}(\tilde{X}_n))$.

Secondly, we construct an interpretable model, typically a linear regression model, to approximate the forecasted behavior of the DNN. The interpretable model $G$ is represented as follows:

$$G(\tilde{X}) = \eta_0 + \sum_{i=1}^{d} \eta_i \tilde{x}_i, \tilde{X} = (\tilde{x}_1, \tilde{x}_2, ... \tilde{x}_d) \tag{14}$$

where $d$ is the number of input features; $\tilde{x}_i$ is the $i^{\text{th}}$ feature of the perturbed instance; and $\eta_i$ is the coefficient of the interpretable model.

Thirdly, the coefficient $\eta_i$ in the interpretable model $G$ is optimized by minimizing the following loss function [22]:

$$L(G) = \sum_{i=1}^{n} \left( \text{DNN}(\tilde{X}_i) - G(\tilde{X}_i) + \Omega(G) \right) \tag{15}$$

where $\Omega(G)$ is a regularization term to control the complexity of the interpretable model.

Finally, the importance of each input feature on the given specific instance is estimated by examining the $\eta_i \tilde{x}_i$ in the interpretable model, since this interpretable model fits this DNN by minimizing the loss function. Larger absolute values of the $\eta_i \tilde{x}_i$ indicate that the corresponding features have a greater influence on the forecasts.

## IV. CASE STUDY

### A. Simulation Settings

#### 1) Dataset Description

To test the newly proposed methods, we will conduct a series of simulations using seven real wind power datasets obtained from three reliable sources, including the JUVENT wind farm in Switzerland, the National Renewable Energy Laboratory (NREL) [23], and GEFCom 2014 [24]. An overview of the dataset characteristics is summarized in Table I for reference.

JUVENT: Datasets from JUVENT are private and require

permission to access. The timeframe ranges from Jan. 1, 2017 to Dec. 31, 2020. Two wind power datasets are randomly selected. One is used to perform wind power forecasting a few hours ahead (from 30 minutes to 3 hours), in which the input data only includes historical wind power. We select data points at 30-minute intervals (i.e., the time resolution is 30 minutes). Another dataset is used for wind power forecasting 24 hours ahead. The consortium for small-scale modeling (COSMO) provides the NWP data, including wind speed at height level 6, wind direction at height level 6, air pressure, Mass fraction of cloud liquid water in air, turbulent kinetic energy, and vertical wind shear. After data preprocessing, the time resolution is 1 hour.

GEFCom 2014 [23]: Datasets from GEFCom 2014 are publicly available. The timeframe ranges from Jan. 1, 2012 to Dec. 1, 2013. We select data points at 60-minute intervals (i.e., the time resolution is 1 hour). GEFCom 2014 includes 10 datasets, but due to space limitations, only 4 datasets are randomly selected for simulation. Look ahead time is 24 hours. NWP data includes wind speed and wind direction at 10 meters and 100 meters (denoted WS10, WD10, WS100, WD100), which will serve as inputs for DNNs.

NREL [24]: Datasets from NREL are publicly available. The timeframe ranges from Jan. 1, 2012 to Dec. 31, 2012. We select data points at 15-minute intervals (i.e., the time resolution is 15 minutes). Due to space limitations, one wind power dataset is randomly selected. Look ahead time ranges from 30 minutes to 4 hours. NWP data is not included in datasets, so the input data only includes historical wind powers.

TABLE I
DATASET DESCRIPTION

| Datasets | Data category | Timeframe | Look ahead time | Input data |
|---|---|---|---|---|
| JUVENT | Private | Jan. 2017 to Dec. 2020 | 0.5 h to 3 h; 24 h | historical wind power; NWP data |
| GEFCom 2014 | Public | Jan. 2012 to Dec. 2013 | 24 h | NWP data |
| NREL | Public | Jan. 2012 to Dec. 2012 | 0.5 h to 3 h | historical wind power |

For wind power forecasting a few hours ahead (from 30 minutes to 3 hours), the length of the input feature is determined by empirical analysis and set to 48 when time resolution is 15 minutes [1]. In simple terms, a window of the past 12 hours of historical data is used to forecast future values. While it is worth noting that hyper-parameter optimization could be explored in future work to fine-tune the lengths of input features for specific cases, the uniformity in using identical input features ensures fairness across all models.

*2) Benchmark Model and Computational Environment*

The focus of this paper is not to develop state-of-the-art DNNs, but to optimize the training process and provide interpretability for different DNNs. Therefore, simulations are performed on widely used DNNs (e.g., MLP, LSTM, GRU, and TCN.) to test whether the proposed method can improve the training process and interpret the results.

Hyper-parameter optimization and cross-validation, as recommended in [20], are performed to determine the most effective parameter configurations for each DNN. As an example, suitable structures and parameters of DNNs for wind power forecasting 30 minutes ahead in the JUVENT dataset can be found in [25].

The described simulation involved the use of different DNNs in a computational environment equipped with Tensorflow 2.0. The computational setup consisted of an Intel(R) Core(TM) i5-10210U CPU running at 1.60 GHz (turbo boost up to 2.11 GHz) and 8 GB of RAM.

*3) Evaluation Metrics*

In the case of each dataset, the first 80% portion is allocated for training, and the subsequent 10% is designated for validation. The remaining 10% serves as the test set. Model performance evaluation is conducted on the test set, utilizing common metrics, such as the coefficient of determination ($R^2$), normalized mean absolute error (NMAE), and normalized root mean square error (NRMSE):

$$R^2 = 1 - \frac{\sum_{k=1}^{K}(\hat{y}_k - y_k)^2}{\sum_{k=1}^{K}(y_k - \bar{y})^2} \quad (16)$$

$$NMAE = \frac{1}{K}\sum_{k=1}^{K}|\hat{y}_k - y_k| \quad (17)$$

$$NRMSE = \sqrt{\frac{1}{K}\sum_{k=1}^{K}(\hat{y}_k - y_k)^2} \quad (18)$$

where $\bar{y}$ is the average value of all data points; $y$ and $\hat{y}$ are the real and forecasted values, respectively; and $K$ is the number of data points. The larger the $R^2$ is, the better the model performance. On the contrary, the smaller NRMSE and NMAE are, the better the model performance.

*4) Simulation Design and Implementation*

The subsequent simulations are organized as follows:

Firstly, in Section B, we will test whether the proposed TriOpts can reduce forecasted errors in different DNNs, datasets, and look-ahead times.

Secondly, we will use a dataset derived from GEFCom 2014 as an example to evaluate the training process, time efficiency, and generalization to different optimization algorithms. Specifically, in Section C, we will compare the training processes of DNNs with and without the TriOpts through visual analysis. In Section D, we will evaluate the time efficiency of DNNs with and without the TriOpts. In Section E, we will examine the generalization of the TriOpts to other optimization algorithms.

Thirdly, in section F, we will use MLP as an example to determine whether the TriOpts can enhance the performance of DNNs in probabilistic forecasting, in addition to the deterministic forecasting discussed in Sections B-E.

Finally, we will consider a dataset derived from GEFCom 2014 and MLP as examples to evaluate the interpretability of the proposed LIME and PFI techniques for wind power forecasting in section G.

### B. Performance Tests in Various DNNs and Datasets

To test whether the TriOpts can reduce forecasted errors, various DNNs are optimized by the Adam algorithm with and without the TriOpts on seven previously described datasets from JUVENT, NREL, and GEFCom 2014. Each DNN is run 30 times independently, and then the forecasted errors of the test set are obtained, as shown in Tables II-IV.

Note that the differences in magnitudes of the metrics before and after implementing the TriOpts in the Tables may not appear significant. This is because all of the metrics have been normalized. In practice, we do not simply compare the magnitudes of these metrics, but rather evaluate the percentage change in these metrics before and after implementing the TriOpts.

From Tables II-IV, we can calculate that the TriOpts show a noteworthy reduction in NRMSE, ranging from 0.48% to 16.24%. Similarly, the TriOpts yield a significant reduction in NMAE, varying from 0.49% to 18.87%. Furthermore, the TriOpts exhibit a remarkable enhancement in $R^2$, with the improvement ratio spanning from 0.31% to 13.83%. These illustrate that the TriOpts can help to improve model generalization and reduce forecasted errors of DNNs.

For example, in the context of wind power forecasting with a 3-hour look-ahead time, as observed in the JUVENT dataset, the TriOpts lead to a reduction in NMSE by 10.25% and a decrease in NMAE by 8.42%, when applying the LSTM model. Besides, the TriOpts contribute to an increase in $R^2$ by 8.13% for the LSTM model.

To visualize the result of wind power forecasting, a wind power generation curve with a timeframe of 2 days is randomly selected from the 1st dataset in the GEFCOM dataset. Then, Fig. 4 presents the forecasted values of different DNNs with and without the TriOpts.

By comparing the forecasted values before and after the application of the TriOpts, it is observed that after applying the TriOpts, the forecasted values generated by the DNNs are closer to the actual values. This indicates that the TriOpts can improve the generalization of the model, and better capture the rapid fluctuations (e.g., troughs) in wind power generation. Equivalent results can also be observed in alternative datasets.

TABLE II
THE FORECASTED ERRORS OF THE DIFFERENT DNNs ON A DATASET DERIVED FROM THE JUVENT

| DNNs | Metrics | Forecasted errors of DNNs **without** the TriOpts | | | | | Forecasted errors of DNNs **with** the TriOpts | | | | |
|---|---|---|---|---|---|---|---|---|---|---|---|
| | | Look ahead time is 0.5 h | Look ahead time is 1 h | Look ahead time is 2 h | Look ahead time is 3 h | Look ahead time is 24 h | Look ahead time is 0.5 h | Look ahead time is 1 h | Look ahead time is 2 h | Look ahead time is 3 h | Look ahead time is 24 h |
| MLP | NRMSE | 0.117 | 0.131 | 0.173 | 0.179 | 0.176 | 0.110 | 0.127 | 0.161 | 0.176 | 0.174 |
| | NMAE | 0.075 | 0.083 | 0.114 | 0.130 | 0.118 | 0.071 | 0.083 | 0.108 | 0.125 | 0.114 |
| | $R^2$ | 0.839 | 0.793 | 0.652 | 0.637 | 0.677 | 0.857 | 0.807 | 0.699 | 0.647 | 0.684 |
| LSTM | NRMSE | 0.113 | 0.128 | 0.159 | 0.175 | 0.178 | 0.111 | 0.123 | 0.143 | 0.147 | 0.173 |
| | NMAE | 0.075 | 0.082 | 0.106 | 0.128 | 0.122 | 0.072 | 0.080 | 0.097 | 0.104 | 0.116 |
| | $R^2$ | 0.850 | 0.804 | 0.705 | 0.651 | 0.668 | 0.855 | 0.818 | 0.763 | 0.741 | 0.687 |
| GRU | NRMSE | 0.114 | 0.128 | 0.157 | 0.162 | 0.179 | 0.110 | 0.125 | 0.143 | 0.146 | 0.174 |
| | NMAE | 0.075 | 0.082 | 0.104 | 0.115 | 0.120 | 0.071 | 0.081 | 0.103 | 0.099 | 0.117 |
| | $R^2$ | 0.848 | 0.804 | 0.713 | 0.686 | 0.664 | 0.858 | 0.812 | 0.769 | 0.760 | 0.682 |
| TCN | NRMSE | 0.137 | 0.148 | 0.169 | 0.178 | 0.178 | 0.125 | 0.140 | 0.146 | 0.156 | 0.170 |
| | NMAE | 0.088 | 0.097 | 0.114 | 0.125 | 0.119 | 0.080 | 0.090 | 0.097 | 0.105 | 0.115 |
| | $R^2$ | 0.780 | 0.736 | 0.669 | 0.639 | 0.668 | 0.817 | 0.765 | 0.753 | 0.716 | 0.697 |

TABLE III
THE FORECASTED ERRORS OF THE DIFFERENT DNNs ON A DATASET DERIVED FROM THE NREL

| DNNs | Metrics | Forecasted errors of DNNs **without** the TriOpts | | | | Forecasted errors of DNNs **with** the TriOpts | | | |
|---|---|---|---|---|---|---|---|---|---|
| | | Look ahead time is 0.5 h | Look ahead time is 1 h | Look ahead time is 2 h | Look ahead time is 3 h | Look ahead time is 0.5 h | Look ahead time is 1 h | Look ahead time is 2 h | Look ahead time is 3 h |
| MLP | NRMSE | 0.077 | 0.092 | 0.142 | 0.178 | 0.075 | 0.089 | 0.138 | 0.174 |
| | NMAE | 0.038 | 0.057 | 0.086 | 0.117 | 0.037 | 0.054 | 0.083 | 0.113 |
| | $R^2$ | 0.964 | 0.948 | 0.880 | 0.810 | 0.967 | 0.953 | 0.886 | 0.819 |
| LSTM | NRMSE | 0.085 | 0.084 | 0.134 | 0.163 | 0.082 | 0.086 | 0.130 | 0.157 |
| | NMAE | 0.051 | 0.049 | 0.080 | 0.104 | 0.046 | 0.047 | 0.077 | 0.098 |
| | $R^2$ | 0.954 | 0.958 | 0.894 | 0.841 | 0.959 | 0.956 | 0.900 | 0.852 |
| GRU | NRMSE | 0.075 | 0.087 | 0.131 | 0.160 | 0.071 | 0.084 | 0.128 | 0.158 |
| | NMAE | 0.036 | 0.049 | 0.076 | 0.104 | 0.032 | 0.045 | 0.077 | 0.098 |
| | $R^2$ | 0.967 | 0.955 | 0.898 | 0.847 | 0.970 | 0.958 | 0.903 | 0.851 |
| TCN | NRMSE | 0.075 | 0.089 | 0.142 | 0.175 | 0.074 | 0.086 | 0.137 | 0.172 |
| | NMAE | 0.039 | 0.052 | 0.082 | 0.120 | 0.038 | 0.047 | 0.080 | 0.108 |
| | $R^2$ | 0.965 | 0.953 | 0.881 | 0.816 | 0.968 | 0.957 | 0.888 | 0.822 |

TABLE IV
THE FORECASTED ERRORS OF THE DIFFERENT DNNs ON DATASET DERIVED FROM THE GEFCOM 2014

| DNNs | Metrics | Forecasted errors of DNNs **without** the TriOpts | | | | Forecasted errors of DNNs **with** the TriOpts | | | |
|---|---|---|---|---|---|---|---|---|---|
| | | 1st dataset in GEFCOM | 2nd dataset in GEFCOM | 3rd dataset in GEFCOM | 4th dataset in GEFCOM | 1st dataset in GEFCOM | 2nd dataset in GEFCOM | 3rd dataset in GEFCOM | 4th dataset in GEFCOM |
| MLP | NRMSE | 0.178 | 0.131 | 0.165 | 0.157 | 0.172 | 0.129 | 0.162 | 0.152 |
| | NMAE | 0.127 | 0.093 | 0.115 | 0.108 | 0.121 | 0.090 | 0.111 | 0.104 |
| | $R^2$ | 0.731 | 0.762 | 0.632 | 0.703 | 0.738 | 0.766 | 0.643 | 0.710 |
| LSTM | NRMSE | 0.175 | 0.132 | 0.169 | 0.157 | 0.171 | 0.130 | 0.163 | 0.154 |
| | NMAE | 0.127 | 0.093 | 0.117 | 0.109 | 0.122 | 0.091 | 0.113 | 0.106 |
| | $R^2$ | 0.732 | 0.759 | 0.629 | 0.703 | 0.736 | 0.763 | 0.633 | 0.708 |
| GRU | NRMSE | 0.178 | 0.131 | 0.172 | 0.156 | 0.172 | 0.130 | 0.164 | 0.151 |
| | NMAE | 0.129 | 0.093 | 0.117 | 0.108 | 0.123 | 0.090 | 0.112 | 0.103 |
| | $R^2$ | 0.731 | 0.762 | 0.627 | 0.705 | 0.736 | 0.767 | 0.634 | 0.711 |
| TCN | NRMSE | 0.176 | 0.130 | 0.169 | 0.157 | 0.172 | 0.128 | 0.164 | 0.152 |
| | NMAE | 0.127 | 0.091 | 0.116 | 0.108 | 0.122 | 0.089 | 0.112 | 0.103 |
| | $R^2$ | 0.732 | 0.766 | 0.630 | 0.703 | 0.737 | 0.771 | 0.635 | 0.710 |

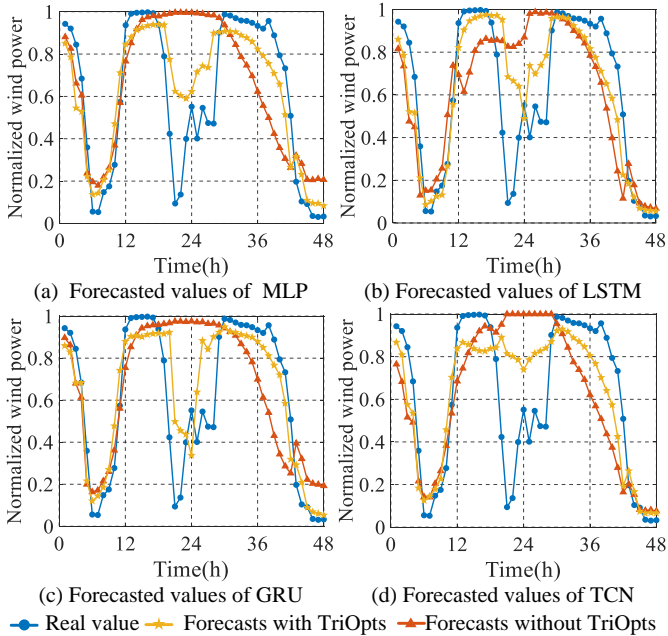

Fig. 4. The real wind power and forecasted wind power of different DNNs with and without the TriOpts.

### C. The Comparison of Training Processes

To compare the training processes of DNNs with and without the TriOpts, the 1st dataset derived from GEFCom 2014 is selected as an example to analyze the training process through visual analysis. Specifically, Fig. 5 presents the training process of different DNNs with and without the TriOpts on this selected wind power dataset.

Obviously, after embedding the TriOpts into the Adam algorithm, significant reductions in both training and validation losses of different DNNs are observed. This observation highlights the potential of the TriOpts to improve model generalization and thereby reduce errors in wind power forecasting.

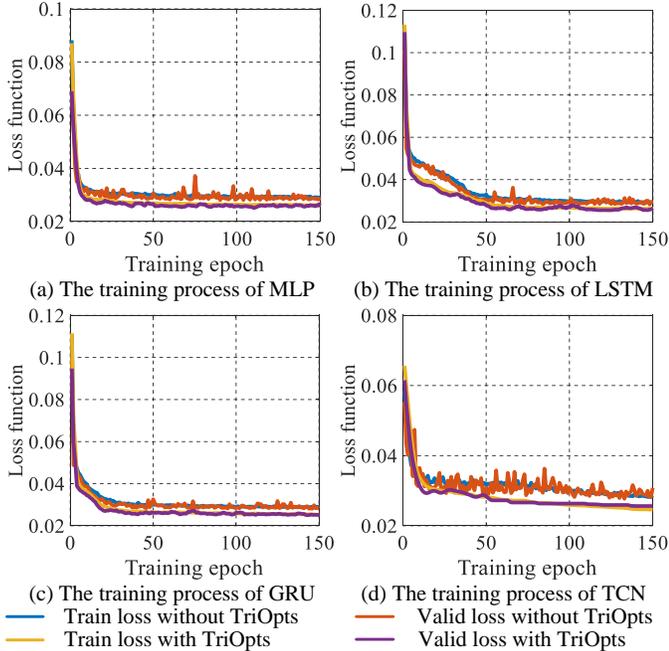

Fig. 5. The training processes of different DNNs with and without the TriOpts.

### D. The Comparison of Time Efficiency

To evaluate the time efficiency of DNNs with and without the TriOpts, the four datasets derived from GEFCom 2014 are selected as an example to analyze the training process through visual analysis. Specifically, Fig. 6 presents the training time of different DNNs with and without the TriOpts on these selected wind power datasets.

A reduction in the training time of DNNs can be observed when the TriOpts are applied to the Adam algorithm, due to the acceleration provided by the proposed gradient centralization strategy. As an illustrative example, the TriOpts lead to a decrease in the training time of MLP by 24.81%, 20.98%, 27.71%, and 14.65% on the first, second, third, and fourth datasets, respectively.

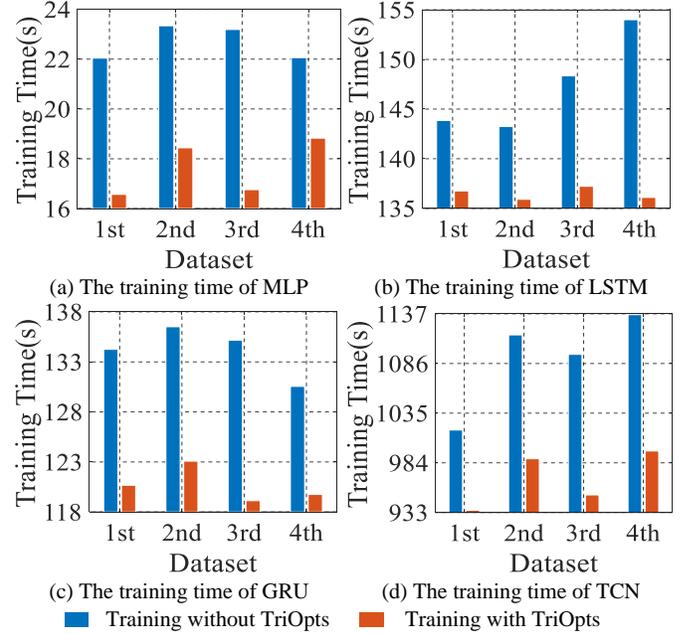

Fig. 6. The training time of different DNNs with and without the TriOpts.

### E. The Generalization to Other Optimization Algorithms

Previous sections have verified the superiority of the TriOpts in one of the most popular and widely used optimization algorithms (i.e., Adam). In this section, we will examine the generalization of the TriOpts to other optimization algorithms, including RMSprop, Nadam, and Adamax [14]. Specifically, the 1st dataset derived from GEFCom 2014 is selected as an example, and then Table V presents the forecasted errors with and without the TriOpts on this selected wind power dataset.

Note that certain optimization algorithms, such as stochastic gradient descent (SGD), adaptive gradient descent (Adagrad), and adaptive delta (Adadelta) in [18], are also tested in our experiments. However, regardless of whether TriOpts are applied or not, the DNNs trained with these algorithms showed significant forecasted errors. This observation underlines their unsuitability for wind power forecasting. Therefore, we will not discuss these algorithms further in this study.

From Table V, it is clear that the TriOpts are not only applicable to the Adam algorithm, but also extend their utility to other optimization algorithms, including Nadam, RMSprop, and Adamax. For example, when the TriOpts are incorporated into the Adamax algorithm, TCN shows a 1.70% and 3.14% reduction in NRMSE and NMAE, respectively, while achieving a 1.51% improvement in $R^2$. This highlights the generalizability of the TriOpts across different optimization algorithms. Compared to the improvements in the Adam algorithm, the improvements for these algorithms seem relatively modest, probably because they are already highly optimized compared to the Adam algorithm, which still has more room for improvement.

TABLE V
THE FORECASTED ERRORS OF DNNs WITH DIFFERENT ALGORITHMS

| DNN | Optimization algorithm | Without the TriOpts | | | With the TriOpts | | |
|---|---|---|---|---|---|---|---|
| | | NRMSE | NMAE | $R^2$ | NRMSE | NMAE | $R^2$ |
| MLP | Nadam | 0.176 | 0.126 | 0.732 | 0.173 | 0.123 | 0.739 |
| | RMSprop | 0.176 | 0.126 | 0.730 | 0.173 | 0.124 | 0.739 |
| | Adamax | 0.178 | 0.126 | 0.725 | 0.174 | 0.124 | 0.736 |
| LSTM | Nadam | 0.176 | 0.125 | 0.731 | 0.173 | 0.123 | 0.738 |
| | RMSprop | 0.176 | 0.128 | 0.731 | 0.173 | 0.125 | 0.738 |
| | Adamax | 0.180 | 0.131 | 0.719 | 0.174 | 0.126 | 0.737 |
| GRU | Nadam | 0.175 | 0.126 | 0.735 | 0.173 | 0.122 | 0.739 |
| | RMSprop | 0.174 | 0.125 | 0.735 | 0.172 | 0.122 | 0.738 |
| | Adamax | 0.176 | 0.128 | 0.730 | 0.173 | 0.125 | 0.738 |
| TCN | Nadam | 0.176 | 0.127 | 0.729 | 0.173 | 0.123 | 0.740 |
| | RMSprop | 0.175 | 0.126 | 0.733 | 0.172 | 0.123 | 0.740 |
| | Adamax | 0.175 | 0.125 | 0.733 | 0.173 | 0.122 | 0.739 |

*F. The Generalization to Probabilistic Forecasting*

Previous sections B-E have verified the superiority of the TriOpts in deterministic wind power forecasting. In this section, we will use MLP with the Adam algorithm and datasets from the GEFCom 2014 as examples to determine whether the TriOpts can enhance the performance of DNNs in probabilistic wind power forecasting.

Specifically, the probabilistic wind power forecasting is realized by the pinball loss function, and the details can be found in [2]. After training the DNNs, the model performance is evaluated by widely used metrics, including quantile score (QS), continuous ranked probability score (CRPS), prediction interval coverage probability (PICP), average coverage error (ACE), prediction interval normalized average width (PINAW), and Winkler score (WS) given a prediction interval nominal confidence(PINC). Their definitions can be found in [1], [2]. The results of probabilistic wind power forecasting with and without the TriOpts are presented in Tables VI and VII.

The TriOpts improve the performance of the probabilistic wind power forecasting, as indicated by the decrease in ACE, PINAW, and WS, and the increase in PICP. A decrease in ACE, which represents the average coverage error, indicates improved accuracy in capturing the true coverage of prediction intervals (PIs). The reduced PINAW indicates narrower PIs. Lower WS values reflect better forecast performance, taking into account both reliability and sharpness. An increased PICP indicates a higher proportion of observations falling within the PIs, demonstrating improved reliability.

For instance, in the 2nd dataset with a 95% PINC, the employment of the TriOpts in the DNN results in a PICP of 0.890, surpassing the PICP of 0.878 observed when the TriOpts are not used. The larger PICP means that the majority of actual wind power values fall within the PIs, demonstrating a higher level of reliability provided by the TriOpts for decision-making in applications, such as power system operation.

Further, the results in Table VI indicate that the proposed TriOpts enhance the performance of probabilistic wind power forecasting. Both the QS and CRPS show reductions when the TriOpts are applied. A decrease in QS implies improved accuracy in forecasting the target quantiles, which means a closer match of forecasts to actual wind power values. Similarly, the decrease in CRPS reflects a reduction in forecasting uncertainty and better calibration of the forecasted distributions, improving the model's ability to provide more reliable and accurate probabilistic wind power forecasting. Considering the accuracy and reliability represented by the QS and CRPS, the DNNs with the TriOpts show much better model performance than those without the TriOpts.

To visualize the result of probabilistic wind power forecasting, a wind power generation curve with a timeframe of 2 days is randomly selected from the 1st dataset in the GEFCOM dataset. Then, Fig. 7 presents the PIs of the MLP with and without the TriOpts given different PINCs. Similar results can also be observed in other datasets and DNNs.

The application of the TriOpts in the MLP results in the production of PIs characterized by a narrower scope, effectively covering the real values of the wind power generation curves. Conversely, the MLP without the TriOpts tends to generate PIs that have a broader and less constrained nature. Too wide PIs can cause wasted capacity in the power system operation. Consequently, the superiority of the TriOpts becomes evident in their improved probabilistic performance.

TABLE VII
THE QS AND CPRS OF DNNs WITH AND WITHOUT TRIOPTS

| Dataset in GEFCOM | Without the TriOpts | | With the TriOpts | |
|---|---|---|---|---|
| | QS | CRPS | QS | CRPS |
| 1st dataset | 0.044 | 0.257 | 0.037 | 0.242 |
| 2nd dataset | 0.040 | 0.181 | 0.033 | 0.171 |
| 3rd dataset | 0.056 | 0.265 | 0.040 | 0.259 |
| 4th dataset | 0.038 | 0.267 | 0.027 | 0.239 |

TABLE VI
THE RESULTS OF PROBABILISTIC WIND POWER FORECASTING WITH AND WITHOUT THE TRIOPTS

| PINC | Metrics | Forecasted errors of DNNs without the TriOpts | | | | Forecasted errors of DNNs with the TriOpts | | | |
|---|---|---|---|---|---|---|---|---|---|
| | | 1st dataset in GEFCOM | 2nd dataset in GEFCOM | 3rd dataset in GEFCOM | 4th dataset in GEFCOM | 1st dataset in GEFCOM | 2nd dataset in GEFCOM | 3rd dataset in GEFCOM | 4th dataset in GEFCOM |
| 80% PINC | PICP | 0.760 | 0.727 | 0.730 | 0.716 | 0.763 | 0.730 | 0.735 | 0.722 |
| | ACE | -0.040 | -0.073 | -0.070 | -0.084 | -0.037 | -0.070 | -0.065 | -0.078 |
| | PINAW | 0.396 | 0.275 | 0.363 | 0.364 | 0.390 | 0.265 | 0.352 | 0.346 |
| | WS | 0.562 | 0.418 | 0.517 | 0.500 | 0.556 | 0.411 | 0.505 | 0.481 |
| 90% PINC | PICP | 0.825 | 0.807 | 0.828 | 0.798 | 0.836 | 0.831 | 0.835 | 0.802 |
| | ACE | -0.075 | -0.093 | -0.072 | -0.102 | -0.064 | -0.069 | -0.065 | -0.098 |
| | PINAW | 0.515 | 0.382 | 0.521 | 0.507 | 0.494 | 0.348 | 0.516 | 0.503 |
| | WS | 0.675 | 0.503 | 0.577 | 0.597 | 0.669 | 0.494 | 0.567 | 0.587 |
| 95% PINC | PICP | 0.891 | 0.878 | 0.854 | 0.830 | 0.895 | 0.890 | 0.858 | 0.833 |
| | ACE | -0.059 | -0.072 | -0.096 | -0.120 | -0.055 | -0.060 | -0.092 | -0.117 |
| | PINAW | 0.721 | 0.547 | 0.583 | 0.612 | 0.685 | 0.481 | 0.573 | 0.583 |
| | WS | 0.815 | 0.623 | 0.636 | 0.689 | 0.784 | 0.576 | 0.626 | 0.665 |

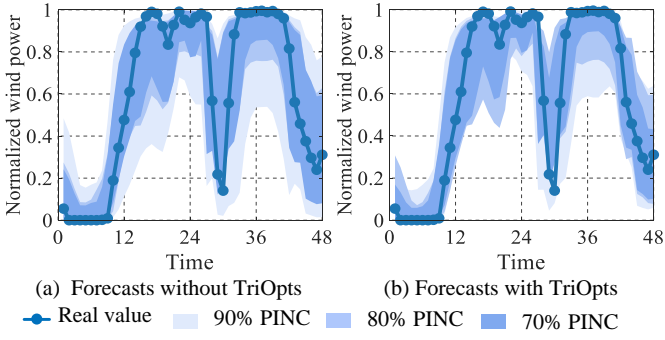

Fig. 7. The PIs of MLP with and without the TriOpts.

### G. Interpretability of Deep Neural Networks

To analyze the interpretability of the proposed LIME and PFI techniques for wind power forecasting, this section will use 1st dataset derived from GEFCom 2014 as an example to interpret the results of wind power forecasting from both the global and instance perspectives.

*1) Global Interpretability*

Global interpretability refers to the calculation of the average contribution (i.e., importance) of each feature in wind power forecasting through the training dataset. Global interpretability has significant implications for feature engineering, as it helps users select relevant features and eliminate unnecessary ones.

Specifically, Fig. 8 presents the normalized contributions of individual features to wind power forecasting in the 1st dataset derived from the GEFCom 2014, in which the features include the wind speeds and wind directions at 10 meters and 100 meters.

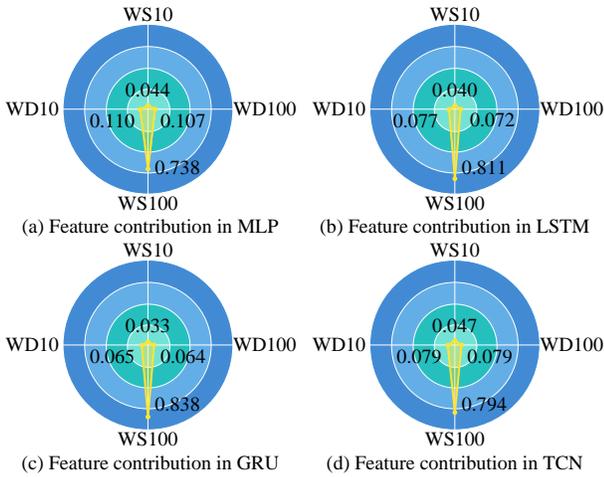

Fig. 8. The average contributions of features (WS10: wind speed at 10 meters; WD10: wind direction at 10 meters; WS100: wind speed at 100 meters; WD100: wind direction at 100 meters).

Compared to the WS10, the WS100 significantly impacts wind power forecasting. This insight aligns with real-world conditions, as wind turbine blades are usually erected at greater heights, meaning wind power production relies heavily on wind speed at higher altitudes (e.g., 100 meters) rather than lower altitudes (e.g., 10 meters). Additionally, WD10 and WD100 have similar contributions for predicting wind power due to consistent wind direction at the 10 and 100-meter levels.

Furthermore, the MLP is used as an example to compare the computational time of the proposed PFI technique with other interpretable techniques (e.g., Shapley values as described in [17]) across various training sample sizes, as shown in Fig. 9.

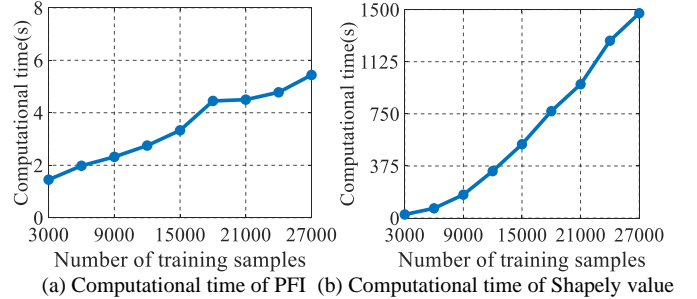

Fig. 9. The computational time of different methods.

From Fig, 9(a), it is observed that the computational time of the proposed PFI technique exhibits an approximately linear relationship with the training sample size, characterized by a relatively slight lope. Even with 27000 training samples, the computational time of the PFI technique remains below 6 seconds. In contrast, as the training sample size increases, the computational time of Shapley values in [17], shows exponential growth, highlighting the significant computational burden imposed by large datasets for Shapley value computations. In comparison to Shapley value-based method in existing publications, the proposed PFI technique demonstrates superior time efficiency.

*2) Instance Interpretability*

Instance interpretability refers to the calculation of the contribution (i.e., importance) of each feature in wind power prediction given a specific sample. Individual interpretability is crucial to comprehending how to obtain forecasted values for a given sample.

Specifically, a sample is randomly selected from the 1st dataset derived from the GEFCom 2014, and then the MLP is used as an example to get the forecasted value of this sample. To interpret how this forecasted value is obtained, the LIME technique is employed to approximate the forecasted behavior of the MLP by constructing an interpretable model, i.e., linear regression in Eq. (14). Note that other DNNs can also be approximated and interpreted in the same way. For this sample, the contribution of each feature to wind power forecasting is shown in Fig. 10.

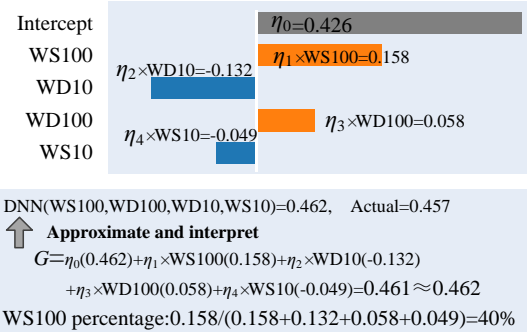

Fig. 10. The contribution of features for a selected sample.

For this selected sample, the forecasted value generated by the MLP is 0.462. However, without the application of the proposed LIME technique, users would remain uninformed about the individual contributions of each feature to wind power forecasting. In this context, the LIME technique employs an interpretable model, such as linear regression, to estimate the forecasted behavior of the MLP, thus elucidating the contributions of each feature for this specific instance.

For example, in this specific sample, the wind speed at 100 meters plays a major role in the forecasts, contributing 0.158 units to the final outcome. Except for the intercept, it contributes 40% to the forecasts. The sum of the contributions (i.e., 0.461) from the four features, along with the intercept, closely approximates the forecasted value of the MLP (i.e., 0.462), demonstrating that the interpretable model effectively fits and explains the forecasted behavior of the MLP. This indicates that the contribution provided by LIME is accurate.

## V. Conclusion

To improve the accuracy and interpretability of DNNs for wind power forecasting, the TriOpts are proposed to accelerate the training process of DNNs, and improve model performance in wind power forecasting. Then, the PFI and LIME techniques are designed to estimate the contribution of individual input features in wind power forecasting, from global and instance perspectives. After conducting simulations on the seven real datasets, the following conclusions can be drawn:

The proposed TriOpts not only improve the model generalization of DNNs for both the deterministic and probabilistic wind power forecasting, but also accelerate the training process. Simulation results show that the TriOpts result in a notable decrease in NRMSE, ranging from 0.48% to 16.24%, and a significant reduction in NMAE, varying from 0.49% to 18.87%. Additionally, they demonstrate a remarkable improvement in $R^2$, with the enhancement ratio spanning from 0.31% to 13.83%. The TriOpts can be used in conjunction with various optimization algorithms (e.g., Adam, Nadam, RMSprop, Adamax, etc.) to enhance their performance.

From a global perspective, the PFI technique can interpret the result of wind power forecasting by estimating the average contribution of each feature, which has significant implications for feature engineering. From an instance perspective, the LIME technique can interpret the result of a given sample by approximating the contribution of each feature to wind power forecasting, which is crucial to understanding how to obtain forecasted values for a given sample.

This paper focuses only on the training process of the deep neural network rather than the design of the structure. Extension work can be done to design more advanced deep neural network structures to improve the model performance. Moreover, the TriOpts and interpretable techniques can be extended to other time series forecasting tasks in future work.